\documentclass[cits]{PoS}

\usepackage{graphicx}
\newcommand{\abb}[1]{Fig.\,\ref{#1}}
\title{The Hydrodynamic Environment for the s Process
in the He-Shell Flash of AGB Stars}

\ShortTitle{Hydrodynamics of He-shell flash convection}

\author{\speaker{Paul R. Woodward}$^{, a}$,
                 David H. Porter$^{a}$,
                 Falk Herwig$^{b,c}$,
                 Marco Pignatari$^{c}$, 
                 Jagan Jayaraj$^{a}$, and 
                 Pei-Hung Lin$^{a}$
                 \\ 
        \llap{$^a$} University of Minnesota, Laboratory for Computational Science \& Engineering
        499 Walter Library, 117 Pleasant St. S. E., Minneapolis, Minnesota 55455, USA\\
        \llap{$^b$}University of Victoria, Dept. of Physics and Astronomy, Victoria, BC, V8W 3P6 Canada\\
        \llap{$^c$} Keele University, Astrophysics Group, North Staffordshire, ST5 5BG, UK\\
        E-mail: \email{paul@lcse.umn.edu, fherwig@uvic.ca}}

      \abstract{The He-shell flash convection in AGB stars is the site
        for the high-temperature component of the s-process in low-
        and intermediate mass giants, driven by the $^{22}\mathrm{Ne}$
        neutron source.  During this phase several s-process
        branchings are activated, including some with time scales
        similar to the convective turn-over time scale.  In addition,
        uncertainties regarding convective mixing at both the bottom
        and the top of the convective shell are preventing accurate
        predictions of s-process yields.  The upper convection
        boundary plays a critical role during the H-ingestion episode
        that may lead to neutron-bursts in the most metal-poor AGB
        stars. We address these problems through global 3-dimensional
        hydrodynamic simulations including the entire spherical
        He-shell flash convection zone (as oposed to the 3D
        box-in-a-star simulations). An important aspect of our current
        effort is to establish the feasibility of our appoach. We
        explain why we favour the explicit treatment over the
        anelastic approximation for this problem. The simulations
        presented in this paper use a Cartesian grid of $512^3$ cells
        and have been run on four 8-core workstations for four days to
        simulate $\sim 5000\mathrm{s}$, which corresponds to almost
        ten convective turn-over times. The convection layer extends
        radially at the simulated point in the flash evolution over
        $7 \mathrm{H_p}$ pressure scale-heights and exceeds the size of
        the underlying core. Convection is dominated by large
        convective cells that fill more than an entire octant. In
        order to better understand the conditions of the s-process
        branchings in this environment we have extracted particle
        tracers, and we discuss the thermodynamic trajectories along
        those paths.}

\FullConference{10th Symposium on Nuclei in the Cosmos\\
		 July 27 - August 1 2008\\
		 Mackinac Island, Michigan, USA}

\begin{document}

\section{Introduction}
The main component of the s process nucleosynthesis originates in AGB
stars, and is intimately linked to the mixing induced by
convection. In the He-shell flash convection zone of stars with
core-masses $>0.6\mathrm{M}_\odot$ 
the $^{22}\mathrm{Ne}(\alpha,\mathrm{n})^{25}\mathrm{Mg}$ reaction
releases a burst of neutrons ($N_\mathrm{n} \sim
10^{11}\mathrm{cm}^{-3}$). This activates s-process branchings
sensitive to the peak temperature reached during the flash, for
example at $^{95}\mathrm{Zr}$ \cite{lugaro:02a} or $^{128}\mathrm{I}$
\cite{reifarth:04}, which in turn depen on the hydrodynamic
properties of this convection, including its boundaries
\cite{herwig:04c}. Observables derived from such branchings make
He-shell flash convection a unique laboratory for investigating
hydrodynamic flows in the deep stellar interior where the elements are
made.

However, our premier motivation are those He-shell flashes that ingest
H from the overlying, unprocessed and stable layers. These have been
found in stellar evolution calculations of AGB stars of extremely low
metal content (e.g. \cite{cassisi:96}), and in other situations
\cite{herwig:01a}.  The convective-reactive nature of the H-ingestion
flash as well as its dependence on the exact nature of convective
boundary mixing, invalidated the one-dimensional stellar evolution
approach, as it for example falsly assumes that the abundance of fuel
is constant on spheres. This assumption may lead to predictions like a
split of convection zones which are most likely artifacts of the
inappropriate assumption of spherical symmetry \cite{woodward:08a}.
Here we focus on the entrainment process in full 3D geomertry that
leads to the convective-reactive case, thereby improving on previous
non-reactive simulations of He-shell flash convection, which where in
plane-parallel geometry and only in two dimensions
\cite{herwig:06a}. First we explain why the explicit hydrodynamic
treatment is justified for this problem.

\section{Hydrodynamic simulations, computational method and
  implications for the s process}

The PPM gas dynamics scheme
\cite{woodward:84,colella:84,woodward:86,woodward:06b} has been
adapted over a decade ago to simulate on 3-D Cartesian meshes AGB star
envelope convection flows in their spherical geometry
\cite{porter:00a,woodward:03}.  In that work the roughly sonic to
supersonic motions near the surface of the star made the explicit
formulation of this numerical scheme highly appropriate. In the
present work, we adopt a similar computational approach to the
simulation of the helium shell flash convection zone in AGB stars.

The principal difficulty introduced through our numerical approach is
our code's explicit formulation, which performs each grid cell update
using only local data.  We follow sound waves in the flow, and these
make the flow tend toward hydrostatic equilibrium
in the radial dimension without our needing to enforce
this constraint explicitly.  This feature of our code is a
disadvantage only because of its computational cost.  Mach numbers in
the helium shell flash convection flow tend to range up to about
1/30\footnote{This is for convective gusts actually observed in our
  simnulations, and not for the much smaller mean convective velocity
  that mixing-length would indicate.}, which means that our explicit
treatment of sound wave propagation forces us to take many more time
steps than would be needed in an anelastic treatment.

\begin{figure}[tbp]
   \includegraphics[width=0.439\textwidth]{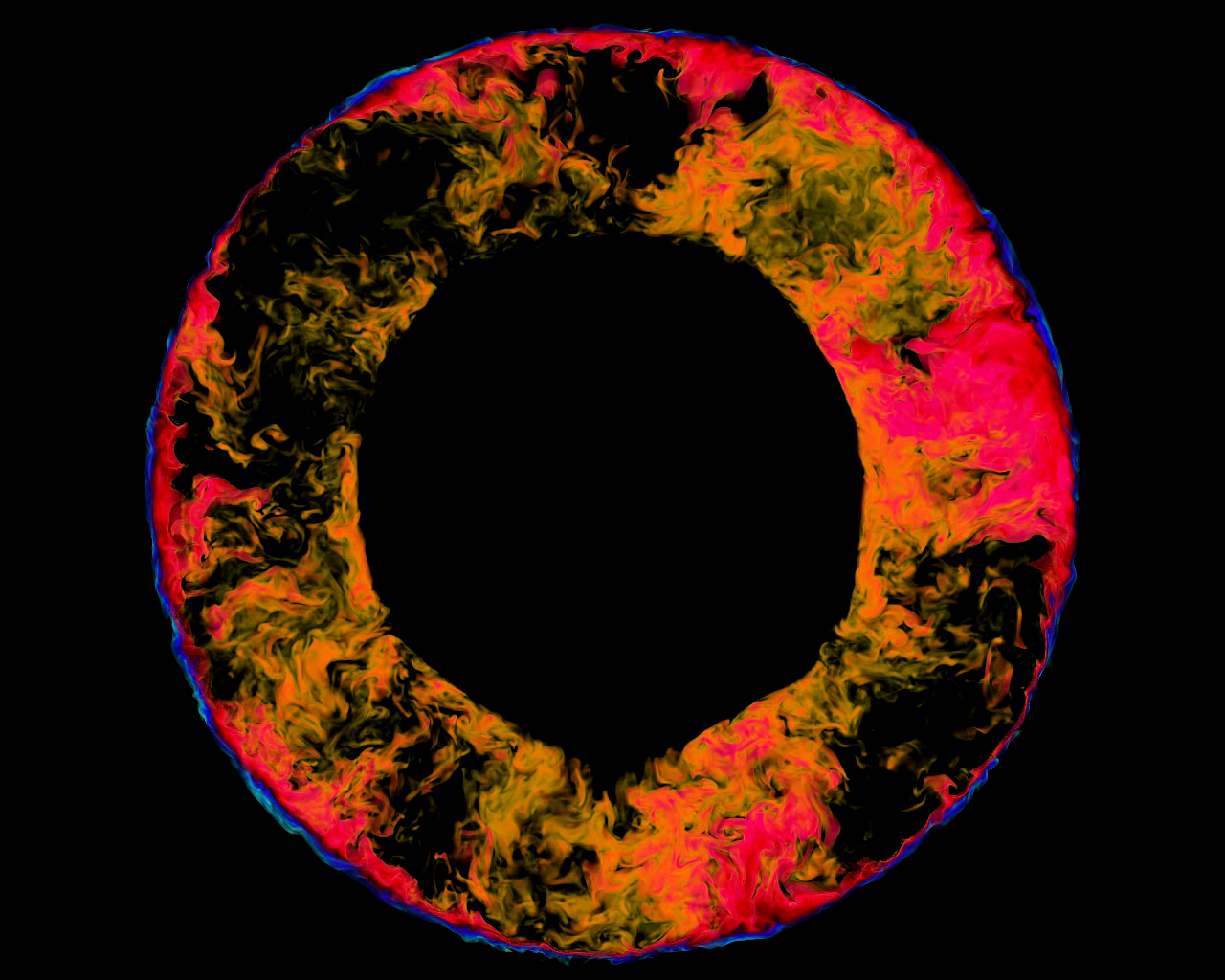} %
   \includegraphics[width=0.561\textwidth]{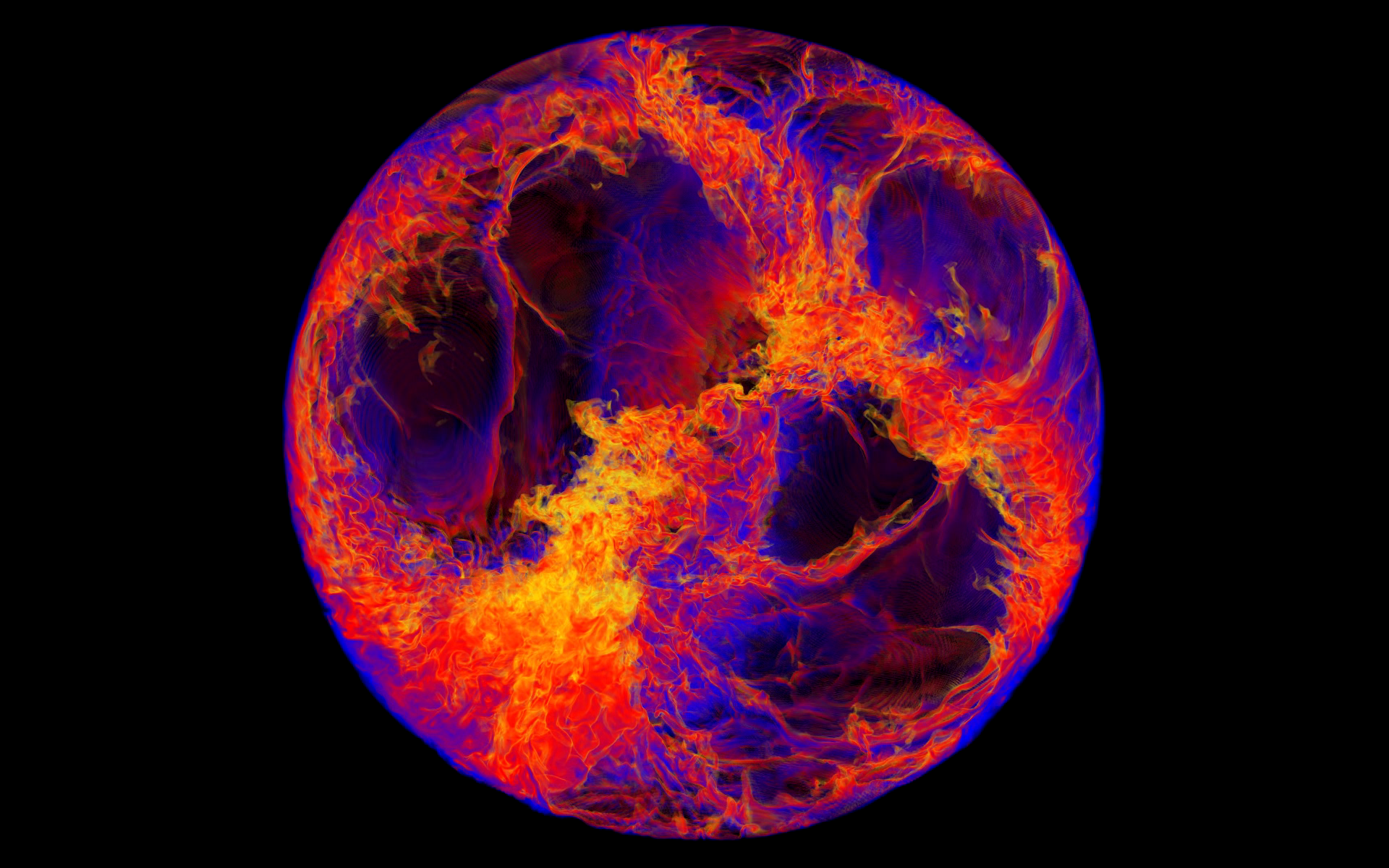} %
   \caption{Snapshots of a hemisphere of the AGB star's interior taken
     from PPM simulations on uniform Cartesian grids of $512^3$ cells,
     corresponding to time $-0.07\mathrm{yr}$ on the left and
     $0.2\mathrm{yr}$ on the right (cf.\ Fig.\, 2 in
     \cite{herwig:06a}).  The outer radius of each image is the upper
     boundary of the convection zone, located on the left at
     $15,200\mathrm{km}$ and on the right at $30,000\mathrm{km}$. The
     bottom of the convection zone is at $9,150\mathrm{km}$ and
     $10,000\mathrm{km}$ in the left and right panel, respectively.
     As before \cite{herwig:06a,woodward:08a} the convection layer is
     sandwiched by a stable layer inside and outside to capture the
     behavior of stable oscillatory motions there that are driven by
     the convection, and artificially damped beyond a certain radius
     in our simulations. Although we do not
     expect the H-ingestion to occur in the high-metallicity AGB star
     that we used as a template for these simulations, we nevertheless
     chose in both cases the entrained material from above the
     convection zone to visualise the flows.  Concentrations of
     entrained fluid from above that are near but not quite equal to
     unity are blue, while red and yellow correspond to concentrations
     of around 1\% and 0.1 to 0.01\%.  Both flows display the dominant
     role of very large convection cells, which tends to invalidate
     the assumptions of the mixing length theory and to support the
     need for detailed treatment of these flows in 3-D in order to
     determine the s-process branching nucleosynthesis.  On the left,
     visualisation focuses on a vertical slice, while on the right the
     hemisphere facing out of the page has been cut away so that we
     look from the inside out at the far hemisphere of the star.
     These images are taken from movie animations that can be viewed
     from the 'MOVIES' link on the main LCSE web site at
     http://www.lcse.umn.edu.}
   \label{fig:3Dfluids}
\end{figure}
However, this numerical approach is affordable because of two aspects.
The first is directional operator splitting, that an anelastic
simulation on the same uniform Cartesian grid could not employ,
because of its much larger time step. This already reduces the the
work difference between the two schemes down by a factor of two to
three. The reason for this is that directional splitting in an
explicit method accurately captures the corner transport for both
sound waves and fluid advection with no additional computational labor
over a 1-D scheme. The second issue is the implicit Poisson solve, or
equivalent, required by the anelastic scheme.  On today's large
machines, global synchronization -- of which many are required per
Poisson solve -- is the most expensive fundamental operation that can
be performed. The second most expensive thing on a modern machine is
communication not overlapped with computation, including -- and not
sufficiently appreciated in the computing community -- a CPU's
communication with its locally attached main memory. The standard
iteration of a Poisson solver offers little computation to perform
while the necessary data is being provided to the CPU by the hardware.
As a result a Poisson solver will run at a fraction of the
Gflop/s/CPU of an explicit gas dynamics scheme like PPM. These factors
together practically eliminate the cost advantage of the anelastic
scheme at Mach numbers seen in helium shell flash convection flows.
In addition, comparisons have shown that perturbations and
instabilities at the boundaries are less prounounced and probably
underestimated in anelastic simulations compared to the explicit
treatment \cite{meakin:07a}. Since the convective boundaries are so
important in our simulations we conclude that the explicit treatment
is the superior method for our problem.

\begin{figure}[tbp]
   \includegraphics[width=0.33\textwidth]{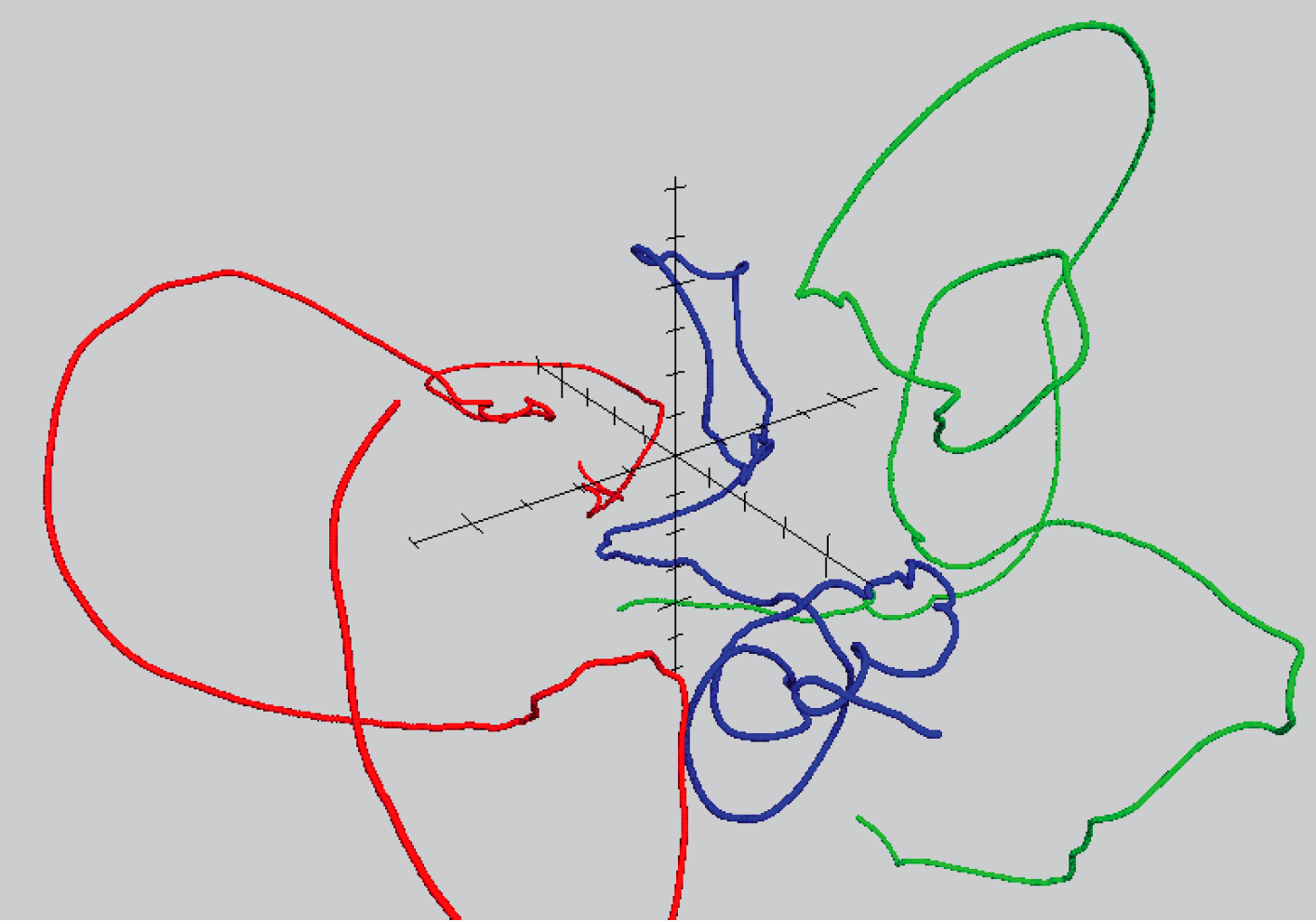}
   \includegraphics[width=0.33\textwidth]{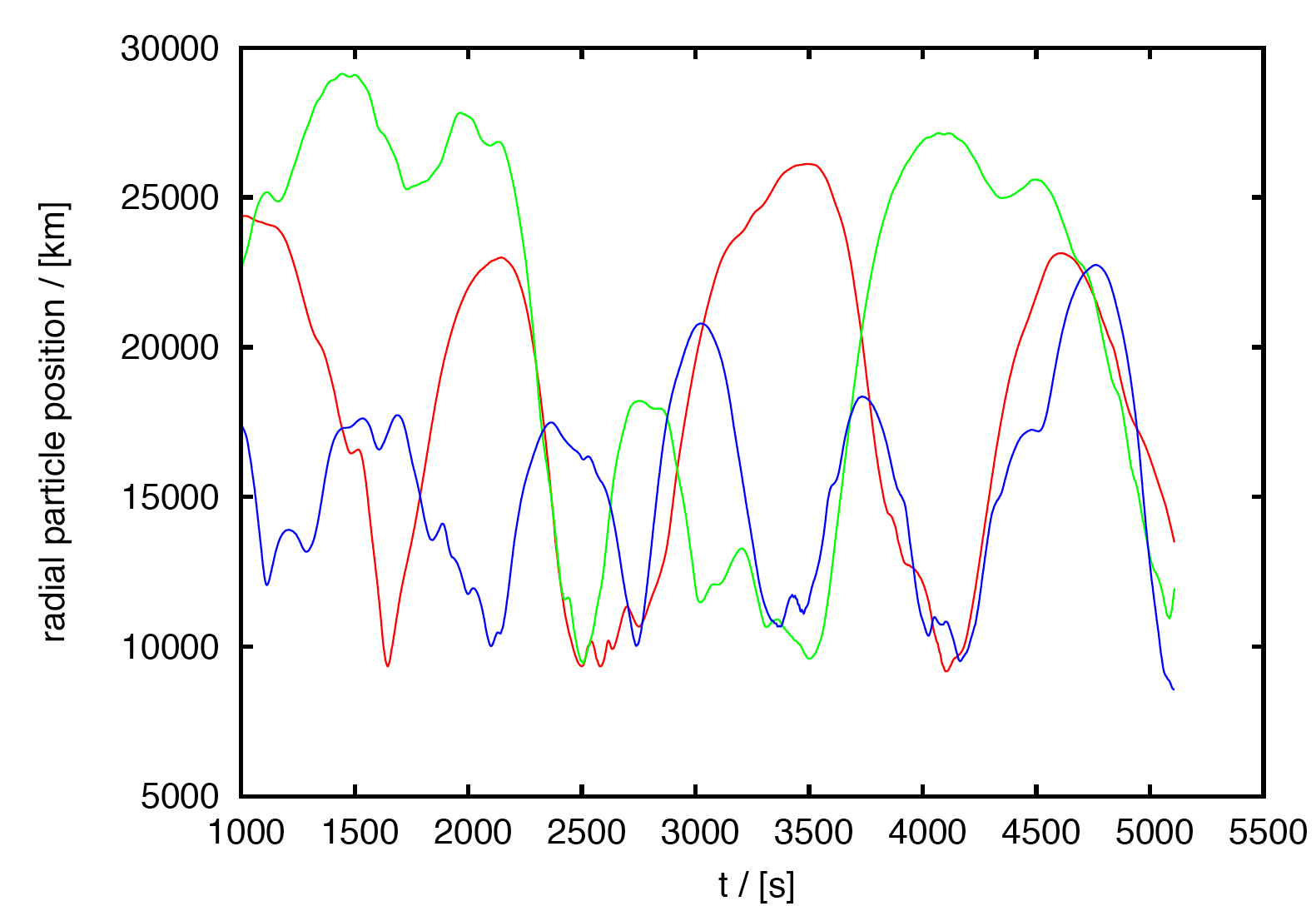}    
   \includegraphics[width=0.33\textwidth]{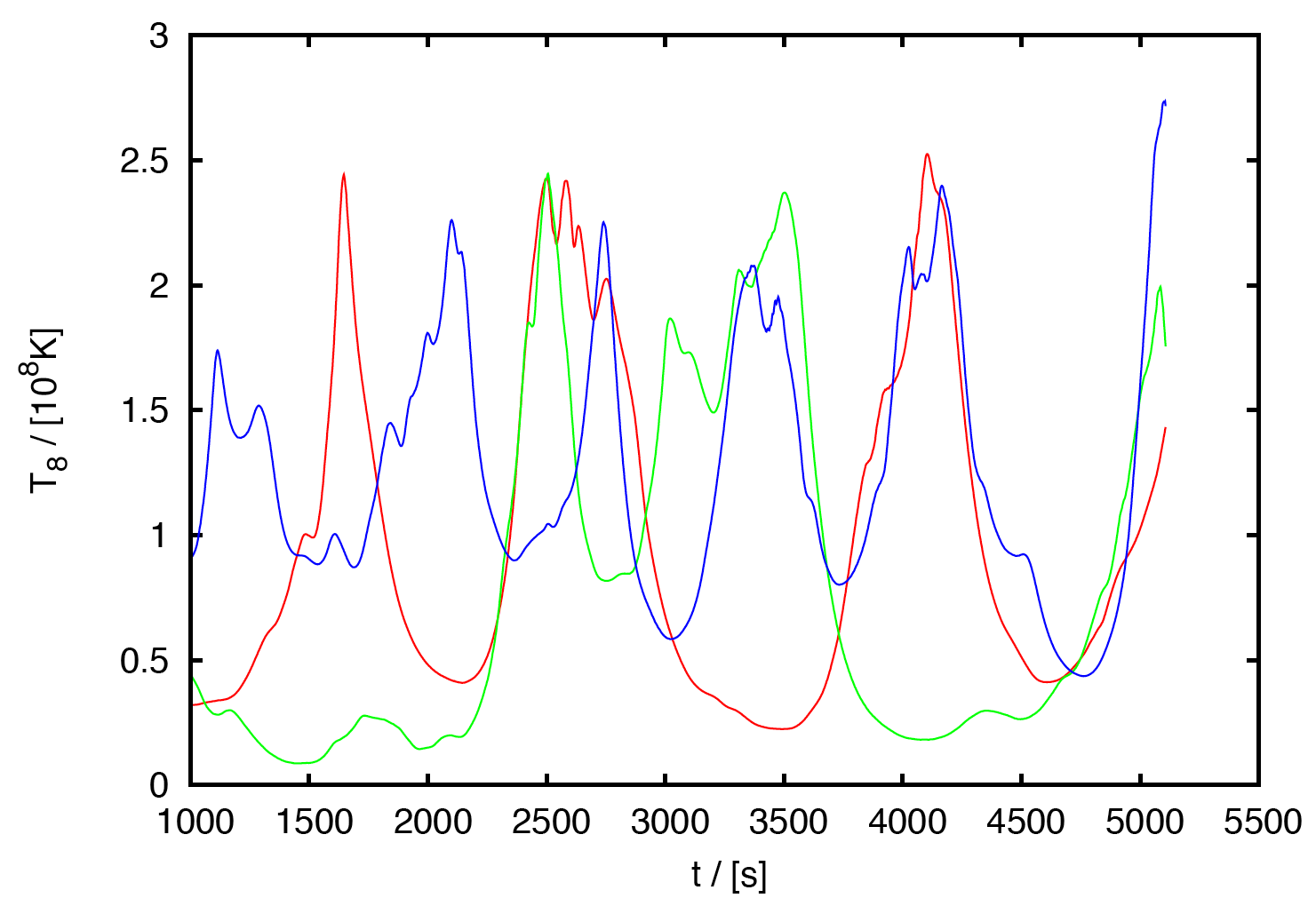} %
   \caption{Left: three arbitrary particle traces over $4176
     \mathrm{s}$ of simulated time in the flow; middle and right: the radial
     position and the temperature of the same particle
     traces.}
   \label{fig:tracers}
\end{figure}


We have implemented our PPM codes on both 4-core Intel CPUs and 8-core
IBM Cell processors (cf. \cite{woodward:08c}). The simulation shown in
the right panel of \abb{fig:3Dfluids} ran over 80,000 time steps to
simulate 5147 seconds of problem time, running for about 4 days on 32
CPU cores in 4 workstations at a sustained performance of 3.4
Gflop/s/core.  An early test of essentially this same two-fluid PPM
code ran at 2.64 Gflop/s/core on 5720 CPU cores (IBM Cell processor
cores) in the Roadrunner machine at Los Alamos, delivering 15 Tflop/s,
which should improve as we perfect our implementation for the
Roadrunner platform.  We expect this code to scale linearly to the
full Roadrunner machine which has over 100,000 CPU cores.
Even if we used only 4\% of that machine we could do this same
simulation on a $1024^3$ grid and with the correct heat injection rate
in just 21 hours, which shows that these computations are eminently
affordable.

In the new global simulations of He-shell flash convection, in
particular at the later time (right panel in \abb{fig:3Dfluids}) when
the geometric extent of the unstable layer is close to the maximum,
the large-scale behaviour of the convection is revealed. Material from
above the convection zone is entrained in a large-scale and irregular
way.  Large cells, that occupy a full quadrant of the thick shell,
dominate the flow, indicating that the mixing patterns relevant for
T-sensitive branchings may be more complicated. 

We can interpolate representative particle paths and the fluid states
along those paths.  Using the fluid states along these particle paths
we can solve a complete nuclear reaction network to obtain the
behavior along this path.  In the simulation, turbulent mixing of
tracked constituents, in this case the original gas of the convection
zone and that entrained from the stably stratified region above it, is
carefully accounted for using a moment-conserving advection scheme
that we call PPB \cite{woodward:86,woodward:08a}.  The PPB scheme has roughly
double or triple the linear resolving power of the PPM advection
scheme, which accounts for the level of fine detail that is evident in
the results shown in \abb{fig:3Dfluids}.  However, the nuclear
reaction network will track a great number of additional
concentrations.  By following many different particle paths
simultaneously, we can use this information to estimate gradients of
these concentrations in a manner similar to the interpolation methods
of smoothed particle hydrodynamics algorithms.  Putting this together
with the PPM simulation's detailed information on the turbulent
kinetic energy along the particle path, an
accurate estimate of s-process yields in this flow can be computed.

\acknowledgments 

This work has been supported by the U. S. Department of Energy through
a contract with the Los Alamos National Laboratory and a grant from
the MICS program of the Office of Science, DE-FG02-03ER25569.  It has
also been supported by NSF Computer Research Infrastructure grant
CNS-0708822, a donation of equipment from IBM, and support from the
Minnesota Supercomputing Institute.  FH was supported in part by Marie
Curie grant MIRG-CT-2006-046520, and JINA.

\providecommand{\bysame}{\leavevmode\hbox to3em{\hrulefill}\thinspace}
\providecommand{\MR}{\relax\ifhmode\unskip\space\fi MR }
\providecommand{\MRhref}[2]{%
  \href{http://www.ams.org/mathscinet-getitem?mr=#1}{#2}
}
\providecommand{\href}[2]{#2}

%

\end{document}